\documentclass[reprint,amsmath,amssymb,aps,pra]{revtex4-1}

\usepackage{graphicx}
\usepackage{bm}
\usepackage{amsthm}
\usepackage{empheq}
\usepackage{rotating}

\usepackage{color}
\newcommand{\Ref}[1]{(\ref{#1})}
\newcommand{\f}{{\bf f}}
\newcommand{\x}{{\bf x}}
\newcommand{\M}{{\bf M}}
\newcommand{\m}{{\bf m}}
 
\newcommand{\R}{\mathbb{R}}

\begin{document}

\preprint{APS/123-QED}

\title{Stabilization of the Lattice Boltzmann Method Using Information Theory}
\author{Tyler L Wilson}
\affiliation{The Fields Institute For Research In Mathematical Sciences}
\author{Mary Pugh}%
\affiliation{%
Department of Mathematics, University of Toronto}%
\author{Francis Dawson}
\affiliation{
 Electrical and Computer Engineering, University of Toronto}%

\date{\today}

\begin{abstract}
A novel Lattice Boltzmann method is derived using the Principle of Minimum Cross Entropy (MinxEnt) via the minimization of Kullback-Leibler Divergence (KLD). By carrying out the actual single step Newton-Raphson minimization (MinxEnt-LBM) a more accurate and stable Lattice Boltzmann Method can be implemented. To demonstrate this, 1D shock tube and 2D lid-driven cavity flow simulations are carried out and compared to Single Relaxation Time LBM, Two Relaxation Time LBM, Multiple Relaxation Time LBM and Eherenfest Step LBM. 
\end{abstract}

\pacs{Valid PACS appear here}

\maketitle

\section{\label{sec:introduction}Introduction}
The Lattice Boltzmann Method (LBM) is a mesoscale discrete velocity model that has become an increasingly popular method for simulating fluid flows, particularly in complex geometries like porous flow (see \cite{Liu2016} for a recent review). It employs a carefully coordinated discretization of physical space, velocity space and time, to track the evolution of a vector valued mass expectation distribution, ${\bf f}$ (whose vector components are sometimes referred to as ``populations.") This evolution is carried out in a cycle of ``streaming'' and ``collision'' steps (see \S \ref{sec:LBM}). 

Historically, the LBM evolved \cite{McNamara1988,Higuera1989,Higuera1989a,Chen1991,Benzi1992,Chen1992,Qian1992} from the Lattice Gas Automata (LGA \cite{Rivet2005,Chopard2005}) to address undesirable features of the LGA such as statistical noise. However, a second interpretation of the LBM is that it is a finite difference form of the continuous Boltzmann Equation \cite{He1997,He1997a,He1997c,Banda2006}. The finite difference view of the LBM allowed researchers to explore various aspects of the method by performing the discretization using various quadratures and lattices.  

Despite this progress it became clear that the LBM can suffer from numerical instabilities \cite{Sterling1996} and so increasing the stability of the LBM became the focus of significant effort. Indeed, a growing number of researchers felt that the lack of unconditional stability (particularly with thermal Lattice Boltzmann Methods) was due to the lack of a so-called ``H-Theorem" for the LBM. Such an H-Theorem would draw inspiration from Boltzmann's H-Theorem for a classical gas \cite{BOLTZMANN2003}.

The first attempts to equip the LBM with an H-Theorem was to 
retain the Single Relaxation Time (SRT-LBM) collision step \Ref{bgkcollision} and replace the Maxwell-Boltzmann \Ref{cnts-MB} based polynomial equilibrium \cite{Qian1992} with an equilibrium that minimizes some entropy function \cite{karlin1998,karlin1998a,renda1998,Frapolli2015,Frapolli2016,Zadehgol2014,Zadehgol2016}. Karlin et al \cite{Karlin1999} proposed that any such entropy function should be convex and, by employing its corresponding minimizer as the equilibrium in the collision step, one should recover the Navier-Stokes equations (up to second order in the macroscopic velocity, ${\bf u}$). Such entropy functions are called ``perfect entropy functions." In addition, Succi \cite{Succi2002} further suggested that the entropy's minimizer should be realizable (bounded between 0 and 1), solvable (expressible as an explicit function of the local macroscopic properties) and lead to Galilean invariant evolution equations. 

An alternative path to equipping the LBM with an H-Theorem would be to find a novel collision step entirely. The most popular of these alternative collision steps is the Entropic Lattice Boltzmann Method (ELBM) which was first described by Karlin et al. in 1999 \cite{Karlin1999,Ansumali2000,Ansumali2002a,Karlin1999,Ansumali2003a} and recently extended to thermal, compressible flows \cite{Frapolli2017EntropicFlows}. More recently a modified collision step that incorporates a local entropic stabilizer parameter, $\lambda$, into the higher order moments was proposed by Karlin, B{\"o}sch and Chikatamarla \cite{Karlin2014,Bosch2017EntropicDynamics}. This entropic stabilizer is locally specified to minimize an entropy function. These models are sometimes called ``KBC" models and have since been explored further \cite{Bosch2015,Dorschner2016,Mattila2015,Dorschner2017TransitionalMethod,Dorschner2017Fluid-StructureMethod,Dorschner2017EntropicDimensions,Flint2017AnMHD}. 

In addition to attempts to equip the LBM with an H-Theorem, other entropic methods have been explored such as non-equilibrium entropy limiters \cite{Brownlee2008}, artificial dissipation \cite{Brownlee2007} and Ehrenfests' coarse-graining \cite{Gorban2006,Brownlee2006} and viscosity filters \cite{Ricot2009}. 

Excellent reviews of the early work on equipping the LBM with an H-Theorem can be found in \cite{Succi2002} as well as \cite{Tosi2008} and of other stabilization methods in \cite{Brownlee2011}.

The assumption that the LBM should (or even {\it could}) incorporate entropic principles is not universally accepted. A wide range of other attempts have been made to improve the stability of the LBM. Chief among them is the Multiple Relaxation Time Lattice Boltzmann Method (MRT-LBM)  \cite{DHumieres1992,DHumieres1994,Lallemand2000,DHumieres2002,McCracken2005,Du2006,Chen2010,Du2012}. 

We share the view that a notion of entropy plays an important role in the stability of the LBM and that entropy violations are a cause of numerical instabilities. Moreover, we consider the fundamental quantity in LBM simulations to be the {\em continuous} mass expectation density, $f$, and we treat the discrete ${\bf f}$ of the LBM to be a particular sampling of it. This allows us to connect the evolution of ${\bf f}$ to information theory. In turn it allows us to employ entropy in the information theoretic sense rather than the thermodynamic sense, avoiding the need for a well defined temperature.

In this paper we describe a novel, third interpretation of the LBM. This new interpretation is based on the Principle of Minimum Discrimination Information (or Minimum Cross Entropy) MinxEnt. We will call this LBM based on MinxEnt, ``MinxEnt-LBM."

This paper is organized in the following way: the LBM, and some of its current variations are described in Section \ref{sec:LBM}. Section \ref{maxent} outlines the foundation of our method, MinxEnt and describes the MinxEnt-LBM framework. Section \ref{sec:numerical} gives some numerical results and Section \ref{conclusions} offers final thoughts and conclusions.

\section{\label{sec:LBM}The Lattice Boltzmann Method}
In kinetic theory \cite{Liboff2003}, the evolution of macroscopic properties often involves understanding the behaviour of a mass expectation density, $f$, (hereafter called a ``distribution"), 
 $$f= f({\bf x},{\bf v},t).$$
The LBM aims to understand $f$ by tracking a related sampled version of it. This section aims to explain how $f$ is sampled via discretization and how its evolution is simulated.

\subsection{\label{sec:LBMsimsec}LBM Simulation Procedure}
Consider the discretization of time by finite time step of length $\delta_t$ such that $t_{n+1}=t_{n}+\delta_t$, $n \in \mathbb{N}$. Further consider a discrete set of $D$ dimensional velocities, 
${\bf V} = \{{\bf v}_1,{\bf v}_2,\dots,{\bf v}_b\ | {\bf v}_{\alpha} \in \mathbb{R}^D\},$ and a discrete set of positions organized into a regular lattice, $\Lambda$. The positions on the lattice are such that ${\bf x}_j \in \Lambda$ if and only if 
${\bf x}_i = \x_j + {\bf v} \, \delta t$ for some ${\bf v} \in {\bf V}$ and ${\bf x}_j \in \Lambda$. 

Having discretized velocity, space and time we define the vector-valued distribution, $\f({\bf x}_j,t_n)$ (denoted $\f$ hereafter), by,

\begin{equation} \label{continuous_to_discrete}
f_{\alpha}({\bf x}_j,t_n)= W_{\alpha} \frac{f({\bf x}_j,{\bf v}_{\alpha},t_n) }{\omega({\bf v}_{\alpha})} \qquad i \in \{1,2,\dots,b \}.
\end{equation}
where $W_{\alpha}$ and $\omega({\bf v_{\alpha}})$ are quadrature weights corresponding to the particular choice of ${\bf V}$. A particular example of $W_{\alpha}$ and $\omega({\bf v}_{\alpha})$ is given in \S \ref{KLDiv}.

Given a distribution $\f$  at time $t_n$  we approximate $\f$ at time $t_{n+1}$ with a two-step process; an instantaneous local ``collision step" followed by a ``streaming step.'' This process can be summarized by the equation,
\begin{equation}
f_{\alpha}( \x_j+ {\bf v}_{\alpha} \, \delta t, t_{n+1})= \Delta_{\alpha} [ {\bf f}({\bf x}_j,t_n)]
\qquad \forall i \in \{1, \dots, b\}
\end{equation}
for some choice of local collision rule ${\bf \Delta}$.

With knowledge of ${\bf f}$, macroscopic quantities at ${\bf x}_j$ are calculated via,
\begin{equation}
\int f({\bf x}_j,{\bf v},t_n) \phi({\bf v}) \, d{\bf v} \approx \sum_{i=1}^b f_{\alpha}({\bf x}_j,t_n) \; \phi({\bf v}_{\alpha}) \label{continuous_moment}
\end{equation}
where $\phi({\bf v})$ is some function of ${\bf v}$. For example, to calculate the local density, $\phi({\bf v}) = 1$, resulting in,
$$\rho({\bf x}_j,t_n) = \sum_{i=1}^b f_{\alpha}({\bf x}_j,t_n)$$
and local momentum, $\phi({\bf v}) = {\bf v}$, resulting in,
$$\rho({\bf x}_j,t_n){\bf u}({\bf x}_j,t_n) = \sum_{\alpha=1}^b f_{\alpha}({\bf x}_j,t_n)\; {\bf v}_{\alpha}.$$
Henceforth, to ease notation we will omit the arguments in the local density, $\rho$ and macroscopic velocity ${\bf u}$.

\subsection{Collision Rules}
\subsubsection{Single Relaxation Time, SRT-LBM}
\label{SRT_CR}
The collision step accounts for changes to the components of the
distribution  arising from collisions between fluid
particles as specified by a collision rule.
 
The most popular collision rule is based the linearization of
the kinetic collision term of the LGA \cite{Higuera1989,Higuera1989a} and further approximation by assuming a single relaxation time $\tau$ \cite{Chen1991}:  
\begin{align}
{\bf \Delta}= \f(\x_j,t_n) + \frac{1}{\tau}\left( \f^{\rm eq}(\x_j,t_n) 
-  \f(\x_j,t_n) \right ).
\label{bgkcollision}
\end{align}
where $\tau$ is some predetermined relaxation time that is related to the fluid viscosity and $\f^{\rm eq}$ is appropriately chosen ``equilibrium'' distribution (see \S \ref{eqmdists}). The relationship between $\tau$ and viscosity is lattice dependent and is shown for a specific lattice in \Ref{visco-tau}.  This single relaxation time approach was made more popular in \cite{Qian1992} and assumed the name Lattice BGK (LBGK) owing to its similarity to the Bhatnagar-Gross-Krook kinetic equation \cite{BGK1954}. For this reason it is common in the literature to refer to LBMs using the single relaxation time collision step as ``LBGK." 

\subsubsection{Multiple Relaxation Time, MRT-LBM}
\label{MRT_CR}

Because of its single adjustable parameter, $\tau$, different fluid properties (such as viscosity and the Prandtl number) cannot be independently specified in SRT-LBM simulations. In an attempt to rectify this issue and improve numerical stability researchers returned to the more general linearized collision term of \cite{Higuera1989,Higuera1989a} which allowed for multiple relaxation times during the collision step
\cite{DHumieres1992,DHumieres1994,Lallemand2000,DHumieres2002,McCracken2005,Du2006,Chen2010,Du2012}. This is accomplished via the collision rule:

\begin{align}
{\bf \Delta} =  \f(\x_j,t_n)
+{\bf T}^{-1} {\bf B} {\bf T} \left(\f^{\rm eq}(\x_j,t_n) 
-  \f(\x_j,t_n) \right )
\label{mrtcollision}
\end{align}
where ${\bf B}$ is a diagonal matrix of relaxation times and ${\bf T}$
is an invertible matrix transforming the vector ${\bf f}$ into a vector of ``moments'' in ``moment space".
Note that the SRT-LBM collision step \Ref{bgkcollision} can be recovered from the MRT-LBM collision by assuming ${\bf B} = \frac{1}{\tau} {\bf I}$. The class of ``Two Relaxation Time'' (TRT-LBM) collision steps was suggested by Ginzburg et al. \cite{Ginzburg2008} and is related to accuracy at boundaries \cite{Ginzbourg1994,Luo2011}. In the TRT-LBM, the diagonal entries of ${\bf B}$ can take only one of two values.  

\subsubsection{SRT-LBM with Ehrenfest Steps (EF-LBM)}
\label{EF_CR}
One attempt to stabilize the LBM using entropic ideas is based on Ehrenfest coarse graining \cite{Ehrenfest1990}. The EF-LBM collision rule equips the LBM with an entropy limiter which monitors
the simulation for lattice points at which some type of local entropic criteria is violated \cite{Brownlee2006}. This approach has been shown to be successful in reducing instabilities in 1-D shock tube  \cite{Brownlee2006,Brownlee2006a,Brownlee2007,Brownlee2007a,Brownlee2008,Brownlee2011}  and 2D lid driven cavity flow \cite{Brownlee2011} simulations.

Given a choice for entropy, $S$, EF-LBM monitors the local nonequilibrium entropy, $\delta S$, defined as, 
\begin{align}
\delta S({\bf f}) := S({\bf f}^{\rm eq})-S({\bf f}). \label{EFlimit}
\end{align}
$\delta S$ serves as an indicator of locations where the non-equilibrium entropy may be too large. If $\delta S$ is below some threshold, a regular SRT-LBM \Ref{bgkcollision} collision step is taken. Otherwise, if the threshold is exceeded, the collision step is altered at that location. For example the collision step in a common version of EF-LBM is,

\begin{subequations}
\label{EFrule}
\begin{empheq}[left={{\bf \Delta} =\empheqlbrace\,}]{align}
\f(\x_j,t_n) +& \frac{1}{\tau}\left( \f^{\rm eq}(\x_j,t_n) 
-  \f(\x_j,t_n) \right ) \nonumber \\
&\mbox{if } \delta S \left ({\bf f} \right) < \mbox{threshold} \\
\f(\x_j,t_n) +& \frac{1}{2\tau}\left( \f^{\rm eq}(\x_j,t_n) 
-  \f(\x_j,t_n) \right ) \nonumber \\
&\mbox{otherwise.} \label{gentle-ef-rule} 
\end{empheq}
\end{subequations}

We can see from \Ref{EFrule} that if $\delta S$ is above the threshold, the approach of EF-LBM is to locally modify the effective relaxation time and reduce the change in $\f$ that occurs during the collision step. This effectively makes the collision step more ``gentle." However, the effect of modifying the effective relaxation time is to locally modify the viscosity.

\subsection{Equilibrium Distributions}
\label{eqmdists}

${\bf f}^{\rm eq}$ is a sampled version of a continuous equilibrium distribution which is to be chosen by the user. For systems involving classical fluid particles, the most widely used equilibrium distribution \cite{Qian1992} is based on the continuous Maxwell-Boltzmann
Distribution:
\begin{align}
f^{\rm MB}({\bf x},{\bf v},t)=\frac{\rho}{(2 \pi R {T})^{D/2}} e^{-\frac{|{\bf v}-{\bf u}|^2}{2R{T}}}.
\label{cnts-MB}
\end{align}

Alternatively, as discussed in the introduction, one path to equipping an LBM with an H-Theorem is to abandon this approach and instead choose an equilibrium that maximizes an entropy function subject to some physical constraints. Explicit examples of  these two cases are given in \S \ref{simsetup}

\section{\label{maxent}The MinxEnt Collision Rule}
\subsection{\label{sec:minxent}Principle of Minimum Cross Entropy (MinxEnt)}
Having given a brief description of the general LBM in \S \ref{sec:LBM}, we now turn to our specific contribution: the MinxEnt collision rule. The MinxEnt collision rule comes from an information theoretic approach which we now discuss.

From kinetic theory, one can show that the mass expectation distribution, $f$, is directly related to the probability, $p$, of finding a particle moving with velocity ${\bf v}$ at location ${\bf x}$ and time $t$. Using the same discretization as in \Ref{continuous_to_discrete}, one finds that, 
\begin{align}
{\bf f}({\bf x}_j,t_n)=\rho({\bf x}_j,t_n) \, {\bf p}({\bf x}_j,t_n).
\label{fandpdef}
\end{align}
With this relationship in mind we take the approach that the fundamental quantity of interest should be these {\it probability} distributions. The question then becomes: during an LBM collision step, how does the pre-collision probability distribution ${\bf p}^{\rm pre}$ change to the post-collision probability distribution ${\bf p}^{\rm post}$? We propose appealing to the Principle of Maximum Entropy (MaxEnt) as described by Jaynes in his seminal paper from 1957 \cite{Jaynes1957}. By adopting this approach we will derive a new LBM collision rule.

In his 1957 paper, Jaynes argues that the probability distribution of an event represents our uncertainty of its outcome. As such we should always assign the probability that incorporates all available knowledge of the event and then maximizes our uncertainty. To do otherwise would introduce bias for which we do not have  evidence to support.  It is this argument that we take to be the foundation of our work on the Lattice Boltzmann Method. 

To quantify our uncertainty it is natural to appeal to the Shannon Entropy. Consider an event with $n$ possible outcomes with discrete probabilities $p_1, p_2,...,p_n$. The Shannon Entropy is given by,
\begin{align}
H(p_{1},p_{2},...,p_{n})=-K\sum_{i=1}^{n}p_{i} \ln p_{i} \label{shanent}
\end{align}
where $K$ is a constant. It was proven by Khinchin in 1957 \cite{Khinchin1957} that \Ref{shanent} was the only function that is
\begin{enumerate}
\item non-negative
\item continuous and symmetric in $p_i$
\item additive for independent sources of uncertainty
\item attains its maximum value when all outcomes are equally likely.
\end{enumerate}
These are the properties that are desirable to have for a function that quantifies uncertainty.

However, even though the LBM tracks discrete distributions $\f$ (equivalently {\bf p} via \Ref{fandpdef}), we must remember that we are treating these as discrete samples of an underlying continuous functions $f$ and $p$. Thus we need a functional that generalizes the Shannon Entropy for situations where the functional acts on continuous probability distributions.

One such functional is the Kullback-Leibler
  Divergence (KLD) \cite{Kullback1951,Ihara1993},
$$
H_{\rm KL} [p | q]= \int_{\R^D} p({\bf v}) \ln \left (\frac{p({\bf v})}{q({\bf v})} \right) \, d {\bf v},
$$
where $q$ is some predetermined reference probability distribution. Note that unlike the Shannon Entropy, the Kullback-Leibler Divergence, $H_{\rm KL}$, does not have a negative sign and so the maximization problem for $H$ becomes a minimization problem for $H_{\rm KL}$. This is called the Principle of Minimum Cross Entropy.

To incorporate our available knowledge of the system we utilize a system of constraints. Let $\{C_1,C_2,...,C_k\}$ be a set of $k$ constraint functionals specific to the physical system of interest, 
$C_i: \mathcal{P} \to \mathbb{R} \qquad i=1,...,k
$
with $\mathcal{P}$ the space of all probability distributions on $\mathbb{R}^D$. Let $\{c_1,c_2,...c_k\}$ be the set of $k$ constraint values. 
Generally speaking the constraint values $c_1,...c_k$ will usually depend on the pre-collision probability distribution  $p^{\rm  pre}$. For example, if momentum is conserved during a collision step we would have $D$ functionals, $C_i(p) = \int {\bf v}_i \, p({\bf v}) \, d{\bf v}$ and $D$ constraint values $c_i = {\bf u}_i$. That is,

$$C_i(p) = c_i \Leftrightarrow  \int {\bf v}_i \, p({\bf v}) \, d{\bf v}  = {\bf u}_i \Leftrightarrow \int {\bf v}_i \, f({\bf v}) \, d{\bf v}  =  \rho{\bf u}_i,$$
where $i \in {1,..,D}$.

Armed with $H_{KL}$ and a set of constraints, the MinxEnt collision rule
is,
\begin{align}\notag
p^{\rm post} =&\arg\!\min_{p\in \gamma} \quad H_{KL}[p|q]\\
\gamma=& \left \{ p \; | \; p \in \mathcal{P}, \;C_1(p) = c_1,...,C_k(p)=c_k \right \} \label{minxentcollision-prob}
\end{align}

\subsection{\label{sec:minxent-collision}MinxEnt-LBM}
The main premise of the MinxEnt-LBM is to apply the Principle of Minimum Cross Entropy as the local collision step within the usual LBM framework (see \S \ref{sec:LBM}). That is, in MinxEnt-LBM the only modification to standard LBM methods will be to choose the post-collision
probability distribution, ${\bf p}^{\rm post}$,  
as the probability distribution that minimizes the Kullback-Leibler Divergence subject to physical constraints. The collision rule will then be constructed from ${\bf p}^{\rm post}$ via \Ref{fandpdef},
$${\bf \Delta}= \rho({\bf x}_j,t_n) \, {\bf p}^{\rm post}({\bf x}_j,t_n).$$

We can discretize the entropy using a given velocity scheme,
\begin{align}
H_{KL}[p| q] &=\int_{\R^D} p({\bf v}) \ln \left ( \frac{p({\bf v})}{q({\bf v})} \right ) \, d {\bf v} \nonumber\\
&\approx \sum_{\alpha=1}^{b} p_{\alpha} \ln \left (\frac{p_{\alpha}}{q_{\alpha}} \right ), \nonumber \\
&:= \mathcal{H}_{\bf q}({\bf p}) \label{discreteEntropy}
\end{align}
where $p({\bf v})$ and ${\bf p}$ are related via \Ref{continuous_to_discrete} and \Ref{fandpdef}, as are $q({\bf v})$ and ${\bf q}$.

This leads us to the general MinxEnt-LBM collision rule,
\begin{align}\notag
{\bf \Delta} &= \rho \, {\bf p}^{\rm post}, \qquad
{\bf p}^{\rm post} =\arg\!\min_{{\bf p}\in \tilde{\gamma}} \quad \mathcal{H}_{\bf q}({\bf p})\\
\tilde{\gamma}&= \left \{ {\bf p} \; | \;  \widetilde{C_1}({\bf p}) = c_1,...,\widetilde{C_k}({\bf p})=c_k \right \} \label{minxentLBM-collision}
\end{align}
where the $\widetilde{C_k}$ are the discretized versions of the corresponding continuous constraint functionals, $C_k$ and ${\bf p}$ is related to $p \in \mathcal{P}$ via \Ref{continuous_to_discrete}.

\section{\label{sec:numerical}Numerical Applications}

To demonstrate the MinxEnt-LBM method, a number of numerical simulations of an athermal 2-D, isotropic Newtonian fluid were carried out.  For comparison, simulations were also carried out with MRT-LBM, SRT-LBM, TRT-LBM and EF-LBM. 
\subsection{Discretization Scheme}
\label{KLDiv}

In our work we use the popular D2Q9 (two dimensional, nine velocity) whose velocities and quadrature weights are described in Table \ref{d2q9table1}. 

\begin{table}[h!]
\centering
\caption{D2Q9 Velocity Scheme:}
\label{d2q9table1}
\begin{tabular}{c||c|c|c|c|c|c|c|c|c}
$\alpha$ & 1 & 2 & 3 & 4& 5& 6 & 7& 8& 9\\
\hline
${\bf v}_{\alpha}$ & (1,0) & (0,1) & (-1,0) & (0,-1) & (1,1) & (-1,1) & (-1,-1)& (1,-1) & (0,0) \\
$W_{\alpha}$ & $\frac{1}{9}$ & $\frac{1}{9}$ & $\frac{1}{9}$& $\frac{1}{9}$ &$\frac{1}{36}$ & $\frac{1}{36}$ & $\frac{1}{36}$& $\frac{1}{36}$& $\frac{4}{9}$\\
\end{tabular}
\end{table}

\subsection{Choice of $q({\bf v})$}
If our system is that of a fluid composed of classical particles then it is reasonable that our choice for $q({\bf v})$ is related to the Maxwell-Boltzmann distribution,  $q({\bf v})= f^{\rm MB}/\rho$.
Sampling this choice of $q({\bf v})$ according to \Ref{continuous_to_discrete} the discritized entropy \Ref{discreteEntropy} becomes,
\begin{align}  
\mathcal{H}({\bf p}) := \sum_{\alpha=1}^{9} p_{\alpha} \ln \left ( \frac{p_{\alpha}}{W_{\alpha}} \right ) 
\label{discrete_KLD}
\end{align}
It should be noted that this form of the discretized entropy is the frequently used form of the discretized entropy used in some LBM simulations (\cite{Ansumali2006,Ansumali2003a,Ansumali2002d,Brownlee2007a,Brownlee2007,Brownlee2008,Brownlee2011} for example).

\subsection{Choice of Constraints}

Because we are considering an athermal, isotropic, Netwonian fluid our system satisfies conservation of mass, conservation of momentum, and the condition of an isotropic/Newtonian fluid. In D2Q9, conservation of mass and momentum yield the numerical constraints,
\begin{align}
\tilde{C_1}({\bf p}) := \sum_{\alpha=1}^9 p_{\alpha} &= 1 :=c_1 \label{discreteprobconst}\\
\tilde{C_4}({\bf p}) :=\sum_{\alpha=1}^9 p_{\alpha} & v_{\alpha,x} = u_x :=c_4,\\ 
\label{discretexmomconst}
\tilde{C_6}({\bf p}) :=\sum_{\alpha=1}^9 p_{\alpha} & v_{\alpha,y} = u_y :=c_6, 
\end{align}

For an isotropic Newtonian fluid we want to ensure the local stress tensor, $\sigma_{jk}$, takes the form, 
\begin{align}
\sigma_{jk}({\bf x},t)=\pi({\bf x},t) \delta_{jk} - 2 \mu \varepsilon_{jk} ({\bf x},t),
\label{newtonian-fluid}
\end{align}
where $\pi$ is the hydrostatic pressure, $\mu$ is the shear viscosity and $\varepsilon_{ij}$ is the strain rate tensor,
\begin{align}
\varepsilon_{jk}=\frac{1}{2}\left [ \frac{\partial u_j}{\partial x_k}+\frac{\partial u_k}{\partial x_j}\right].
\label{strain-rate-tensor}
\end{align}

In terms of the mass expectation distribution, this constraint becomes \cite{Liboff2003},
\begin{align}
\int f({\bf x},{\bf v},t) [{\bf v}-{\bf u}]_j & [{\bf v}-{\bf u}]_k \, d{\bf v} \nonumber\\
& = \pi({\bf x},t) \delta_{jk} - 2 \mu \varepsilon_{jk} ({\bf x},t) \label{viscoconstraint}.
\end{align}

Discretized in the D2Q9 scheme these constraints become,
\begin{align}
\tilde{C_8}({\bf p}) := \sum_{\alpha=1}^9 p_{\alpha}& \left (v_{\alpha,x}^2-v_{\alpha,y}^2 \right) \nonumber\\
&=\sum_{\alpha=1}^{9} \left [ p_{\alpha}^{\rm pre} + \frac{1}{\tau}( p_{\alpha}^{\rm eq}-p_{\alpha}^{\rm pre}) \right ] \left (v_{\alpha,x}^2-v_{\alpha,y}^2 \right):=c_8, \label{discreteviscoconst}\\
\tilde{C_9}({\bf p}) := \sum_{\alpha=1}^9 p_{\alpha}& v_{\alpha,x}v_{\alpha,y} \nonumber\\
&=\sum_{\alpha=1}^{9} \left [ p_{\alpha}^{\rm pre} + \frac{1}{\tau}( p_{\alpha}^{\rm eq}-p_{\alpha}^{\rm pre}) \right ] \left (v_{\alpha,x}v_{\alpha,y} \right):=c_9, \label{discreteisoconst}
\end{align}
One can arrive at the numerical constraints \Ref{discreteviscoconst} and \Ref{discreteisoconst} by assuming the equilibrium takes the form \Ref{numerical-plyeqm} and using the Chapman-Enskog expansion \cite{Wilson2016}. The Chapman-Enskog expansion also provides the relationship between $\tau$ and the viscosity $\nu$, as well as pressure $\pi$ and density,

\begin{align}
\nu=\frac{2 \tau -1}{6}\frac{\delta_x^2}{\delta_t}, \qquad \pi= \frac{\rho}{3} \label{visco-tau}
\end{align}

Additional, numerically motivated, constraints can be added to deal with inaccuracies at the boundary \cite{Ginzbourg1994,Luo2011},
 \begin{align}
\tilde{C_5}({\bf p}) :=\sum_{\alpha=1}^9 &p_{\alpha}  \left[-5+3 \left(v_{\alpha,x}^2+v_{\alpha,y}^2 \right) \right] v_{\alpha,x} \nonumber\\
&= \sum_{\alpha=1}^{9} \left [ p_{\alpha}^{\rm pre} + \frac{1}{\tau_2}( p_{\alpha}^{\rm eq}-p_{\alpha}^{\rm pre}) \right ], \nonumber \\
&\qquad \qquad \left[-5+3 \left(v_{\alpha,x}^2+v_{\alpha,y}^2 \right) \right] v_{\alpha,x}:=c_5 , \label{extraconstraint1}\\
\tilde{C_7}({\bf p}) :=\sum_{\alpha=1}^9 &p_{\alpha}  \left[-5+3 \left(v_{\alpha,x}^2+v_{\alpha,y}^2 \right) \right] v_{\alpha,y} \nonumber\\
&= \sum_{\alpha=1}^{9} \left [ p_{\alpha}^{\rm pre} + \frac{1}{\tau_2}( p_{\alpha}^{\rm eq}-p_{\alpha}^{\rm pre} ) \right ] \nonumber\\
&\qquad \qquad \left[-5+3 \left(v_{\alpha,x}^2+v_{\alpha,y}^2 \right) \right] v_{\alpha,y}:=c_7,  \label{extraconstraint2}
\end{align}
where,
  
\begin{align}
\tau_2 &= \frac{8 \tau - 1}{2 \tau -1} \label{tau2}.
\end{align}
This is in the same spirit as TRT-LBM \cite{Ginzburg2005,Ginzburg2008,Ginzburg2005a}. These constraints are purely numerical to increase accuracy of the simulation near the boundaries. 
  
To summarize, in the D2Q9 scheme, the MinxEnt-LBM collision for an athermal 2D isotropic Newtonian fluid becomes the constrained optimization problem:
\begin{align}\notag
{\bf p}^{\rm post} =\arg\!\min_{{\bf p}} \quad \sum_{\alpha=1}^{9} p_{\alpha} \ln \left ( \frac{p_{\alpha}}{W_{\alpha}} \right )
\end{align}
 subject to the constraints \Ref{discreteprobconst}, \Ref{discretexmomconst}, \Ref{discreteviscoconst}, \Ref{discreteisoconst}.  We call this version of MinxEnt-LBM with  5 constraints ``MinxEnt4,'' owing to the 4 free parameters remaining to minimize over. We call the version of MinxEnt-LBM with the 7 constraints, \Ref{discreteprobconst}, \Ref{discretexmomconst}, \Ref{discreteviscoconst}, \Ref{discreteisoconst}, \Ref{extraconstraint1} and \Ref{extraconstraint2}  ``MinxEnt2".

\subsection{Minimization Procedure: MinxEnt-LBM Using Newton-Raphson in Moment Space}
\label{minxent-NR}
In this work we will choose to minimize the entropy using the Newton-Raphson minimization procedure. Before proceeding however we find it convenient to turn the constrained minimization into an unconstrained minimization by moving into moment space using the invertible matrix ${\bf T}$: ${\bf M}:={\bf T}{\bf p}$. Given the form of the constraints and velocities from Table \ref{d2q9table1} it is convenient to define some rows of ${\bf T}$ to correspond to the discrete constraints and thus we choose,
\begin{align}
{\bf T}=  \left[ \begin{array}{ccccccccc}
1 & 1 & 1 & 1 & 1 & 1 & 1 & 1 & 1 \\
-4 & -1 & -1 & -1 & -1 & 2 & 2 & 2 & 2 \\
4 & -2 & -2 & -2 & -2 & 1 & 1 & 1 & 1 \\
0 & 1 & 0 & -1 & 0 & 1 & -1 & -1 & 1 \\
0 & -2 & 0 & 2 & 0 & 1 & -1 & -1 & 1 \\
0 & 0 & 1 & 0 & -1 & 1 & 1 & -1 & -1 \\
0 & 0 & -2 & 0 & 2 & 1 & 1 & -1 & -1 \\
0 & 1 & -1 & 1 & -1 & 0 & 0 & 0 & 0 \\
0 & 0 & 0 & 0 & 0 & 1 & -1 & 1 & -1 
\end{array} \right], \label{Tmatrix}
\end{align}

(The ordering of these moments is taken to be consistent with the literature \cite{Lallemand2000}). 
Choosing ${\bf T}$ with these properties renders the constraints into a simpler form,
\begin{align}
M_1 &=1 , \label{momentconstraint1}\\
M_4 &=u_x,\label{momentconstraint4}\\
M_5 &= M_5^{\rm pre} + \frac{1}{\tau_2}( M_5^{\rm eq}-M_5^{\rm pre} ),\label{momentconstraint5}\\
M_6 &= u_y,\label{momentconstraint6} \\
M_7 &= M_7^{\rm pre} + \frac{1}{\tau_2}( M_7^{\rm eq}-M_7^{\rm pre} ) \label{momentconstraint7}\\
M_8 &= M_8^{\rm pre} + \frac{1}{\tau}( M_8^{\rm eq}-M_8^{\rm pre} )\label{momentconstraint8}\\
M_9 &= M_9^{\rm pre}+ \frac{1}{\tau}( M_9^{\rm eq}-M_9^{\rm pre}). \label{momentconstraint9}
\end{align}

The second and third rows of ${\bf T}$ remain to be chosen. In principle they are arbitrary, provided that${\bf T}$ is invertible. To be consistent with the literature We will use \Ref{Tmatrix}.

Defining the vector of free, unconstrained moments by ${\bf m}$,

the {\it full} vector of moments (including constraints)  as
$${\bf M}=\begin{cases} \langle 1,m_1,m_2, u_x,m_3, u_y,m_4, c_8, c_9 \rangle \qquad &\mbox{MinxEnt4},\\ \langle 1,m_1,m_2, u_x, c_5, u_y, c_7, c_8, c_9 \rangle \qquad &\mbox{MinxEnt2}.\end{cases}$$

To fulfill the MinxEnt-LBM collision step we now seek the vector $\m$ that minimizes the discretized entropy, \Ref{discrete_KLD}, rewritten in moment space:
\begin{align}
\mathcal{S}(\M):&=\mathcal{H}({\bf T}^{-1}{\bf M})\nonumber \\
&=\sum_{\alpha=1}^{9} ( {\bf T}^{-1} {\bf M} )_{\alpha} \, \ln \left ( \frac{ ( {\bf T}^{-1} {\bf M} )_{\alpha}}{W_{\alpha}} \right ). \label{KL-moment} 
\end{align}

Depending on the version of MinxEnt-LBM either five or seven components of $\M$ are fixed by the constraints, and so the constrained minimization problem involves the gradient of $\mathcal{S}$ with respect to the four or two unconstrained moments respectively, 
\begin{align*}
\nabla_k \mathcal{S}(\m)  &= \frac{ \partial \mathcal{S}}{\partial m_k}(\M)\\
&= \sum_{\alpha=1}^{9} ({\bf T}^{-1})_{\alpha k} \left [ \ln \left ( \frac{ ( {\bf T}^{-1}{\bf M})_{\alpha}}{W_{\alpha}}\right ) +1 \right ]
\\
& \qquad \qquad k \in \{1,...,4\} \, \mbox{ or }\{1,2\}.
\end{align*}

The Hessian is
\begin{align*}
{\bf H}_{jk} (\m)&= \frac{ \partial^2 \mathcal{S}}{\partial m_j \partial m_k}(\M)\\ 
&= \sum_{\alpha=1}^{9} \frac{({\bf T}^{-1})_{\alpha k} ({\bf T}^{-1})_{\alpha j}}{({\bf T}^{-1} {\bf M} )_{\alpha}} \\
& \qquad \qquad j,k \in \{1,...,4\} \, \mbox{ or }\{1,2\}.
\end{align*}

We then perform the Newton-Raphson procedure,
\begin{align*}
\m^{n+1}=\m^0-{\bf H}^{-1}(\m^n)\, \nabla \mathcal{S}(\m^n)
\end{align*}
 
For the initial moments in the Newton-Raphson procedure, ${\bf m}^0$ we compute the moments of the discretized equilibrium distribution, ${\bf M}^{0}={\bf T} {\bf p}^{\rm eq}$
 and then take,

 $${\bf m}^0=\begin{cases} \langle M^0_{2},M^0_{3},M^0_{5},M^0_{7} \rangle \qquad &\mbox{MinxEnt4},\\
\langle M^0_{2},M^0_{3} \rangle  \qquad &\mbox{MinxEnt2}.\end{cases}$$

In principle the Newton-Raphson procedure should be continued until some convergence criteria is satisfied, however to reduce computational overhead we terminate after a single step.

This MinxEnt-LBM collision step can be summarized by the following algorithm,
 \begin{enumerate}
 \item Calculate pre-collision moments ${\bf M}^{0} ={\bf T} {\bf p}^{\rm eq}$ and constraints 
 \item Calculate ${\bf H}$ and gradient vector $\nabla \mathcal{S}$ 
 \item Perform a single Newton-Raphson step for the unconstrained moments, 

$$\m^{1}=\m^0-{\bf H}^{-1}(\m^0)\, \nabla \mathcal{S}(\m^0)$$
 \item Construct the full post-collision moment vector, ${\bf M}^{\rm post}$
 \item Return to distribution space ${\bf p}^{\rm post} = {\bf T}^{-1} {\bf M}^{\rm post}$
 \end{enumerate}

\subsection{General Simulation Setup}
\label{simsetup}
In D2Q9 the Maxwell-Boltzmann based polynomial equilibrium is given by,
\begin{align}
f_{\alpha}^{\rm eq}  &=W_{\alpha} \rho  \left \{ 1 + 3{\bf v}_{\alpha}\cdot {\bf u}+ \frac92({\bf v}_{\alpha}\cdot {\bf u})^2 - \frac32 \left |{\bf u}\right|^2 \right \}.\label{numerical-plyeqm}
\end{align}
All LBMs simulated here will use this equilibrium with the exception of EF-LBM.

EF-LBM simulations adopt the entropy minimization approach to the choice of equilibrium, taking ${\bf f}^{\rm eq}$ to minimize the discrete entropy function \Ref{discrete_KLD} subject to physical constraints \Ref{discreteprobconst},\Ref{discretexmomconst}, 
arriving at \cite{Ansumali2003a},
\begin{align}
f_{\alpha}^{\rm eq} &=W_{\alpha} \rho \prod_{j=1}^{2} \left (2 - \sqrt{1 + 3u_j^2} \right ) \left ( \frac{2u_j + \sqrt{1+3u_j^2}}{1-u_j}\right )^{v_{\alpha,j}}. \label{numerical-enteqm} 
\end{align}
The MRT-LBM, TRT-LBM, SRT-LBM and EF-LBM collisions are all based on the same rule \Ref{mrtcollision}.
The only difference between the methods is the choice of the matrix ${\bf B}$. The different versions of ${\bf B}$ are shown in Table \ref{table:relaxation-times}. We will utilize the same matrix ${\bf T}$,  \Ref{Tmatrix}, for all simulations including MinxEnt-LBM. 
 
\begin{table}[h!]
\begin{center}
\begin{tabular}{|c|c|}
\hline
Collision & ${\bf B}$\\
\hline
MRT-LBM &
$ diag\left (0,1.64,1.54,0,\frac{1}{\tau_2},0,\frac{1}{\tau_2},\frac{1}{\tau},\frac{1}{\tau} \right )$\\
\hline
TRT-LBM & $diag\left (0,\frac{1}{\tau},\frac{1}{\tau},0,\frac{1}{\tau_2},0,\frac{1}{\tau_2},\frac{1}{\tau},\frac{1}{\tau} \right )$ \\
\hline
SRT-LBM &$\frac{1}{\tau} {\bf I}$\\
\hline
EF-LBM & $\begin{cases} \frac{1}{\tau} {\bf I} &\mbox{if } \delta S < \mbox{tolerance} \\
\frac{1}{2 \tau} {\bf I}& \mbox{if otherwise.}  \end{cases}$\\
\hline
\end{tabular}
\end{center}
\caption{Relaxation times for the various LBM collisions. The values 1.64 and 1.54 are chosen to agree with \cite{Luo2011,Lallemand2000}. Tolerance values are given in \Ref{tolerances}. $\delta S$ is defined in \Ref{EFlimit}}
\label{table:relaxation-times}
\end{table}

We denote three different versions of EF-LBM by EF1, EF2 and EF3 according to their tolerance values, 

\begin{align}
\mbox{tolerance} = \begin{cases}
\infty &\mbox{if EF1}\\ 10^{-3} &\mbox{if EF2}, \\
10^{-5} &\mbox{if EF3}. \end{cases}
\label{tolerances}
\end{align}

Note that although SRT-LBM and EF1 have the same $\frac{1}{\tau}$ timescale, they are different schemes because of they use different equilibriums; \Ref{numerical-plyeqm} and \Ref{numerical-enteqm} respectively.

All no slip boundary, zero velocity conditions are realized by using the full-way bounceback scheme. For distributions on fixed non-zero velocity boundaries, the components of ${\bf f}$ are assigned the equilibrium distribution according to the macroscopic conditions (density and velocity) required at the boundary.

The initial distributions are set to the equilibrium distribution. 

\subsection{\label{shockstudies}1D Shock tube (ST)}

A benchmark simulation to test the stability of a simulation is the low viscosity 1D shock tube \cite{Sod1978ALaws}. Simulations were performed with values of $\tau$ close to $0.5$ since the zero viscosity limit occurs when $\tau=0.5$, see \Ref{visco-tau}.  The initial condition is chosen to have a density shock located at the centre of the tube.  

The simulation was carried out in a 2D geometry because simulations carried out in 1D use the 1DQ3 scheme. The MinxEnt-LBM method would be over constrained and minimization would not be required. To perform MinxEnt-LBM in a 1D geometry with more than three velocities would require a non-uniform spacing of lattice points \cite{Chikatamarla2006a} and destroy its lattice structure. Thus a 2D simulation with periodic boundary conditions was employed with initial data that was taken to be constant in the $y$ direction.

The remainder of the simulation setup is summarized in the first row of Table \ref{table:shocksetup}.

\begin{table}[h!]
\centering
\begin{tabular}{|c|c|c|c|c|}
\hline
Sim & $(N_x,N_y)$ & IC & SC & $\tau$\\
\hline
ST& (800,4) & $\rho =  1, x \in [0,400]$ & $t_n=400$ & $0.5 + 10^{-9}$\\
\S F&  & $ \rho= .5, x \in [401,800]$ & & \\
&  & ${\bf u}={\bf 0}$ & & \\
\hline
LDSS& (17,17) & $\rho =  2.7, {\bf u}=0 $ & $t_n=1000$ & varied \\
\S G&  &$u_{lid}$=varied  & or fail &  \\
\hline
LDAS& (257,257) & $\rho =  2.7, {\bf u}=0 $ & $\Delta \psi_{min} < $ & varied \\
\S H&&$u_{lid}$=0.01 or  &$10^{-5}$ &  \\
&  & 0.1 & &  \\
\hline
\end{tabular}
\caption{Setup for the simulations. Sim: Simulation Type, $N_x,N_y$: Number of lattice nodes in respective directions, IC: Initial Condition, SC: Stopping Condition}
\label{table:shocksetup}
\end{table}

 \begin{figure}[t!]
        \centering
      \includegraphics[width=.95\columnwidth]{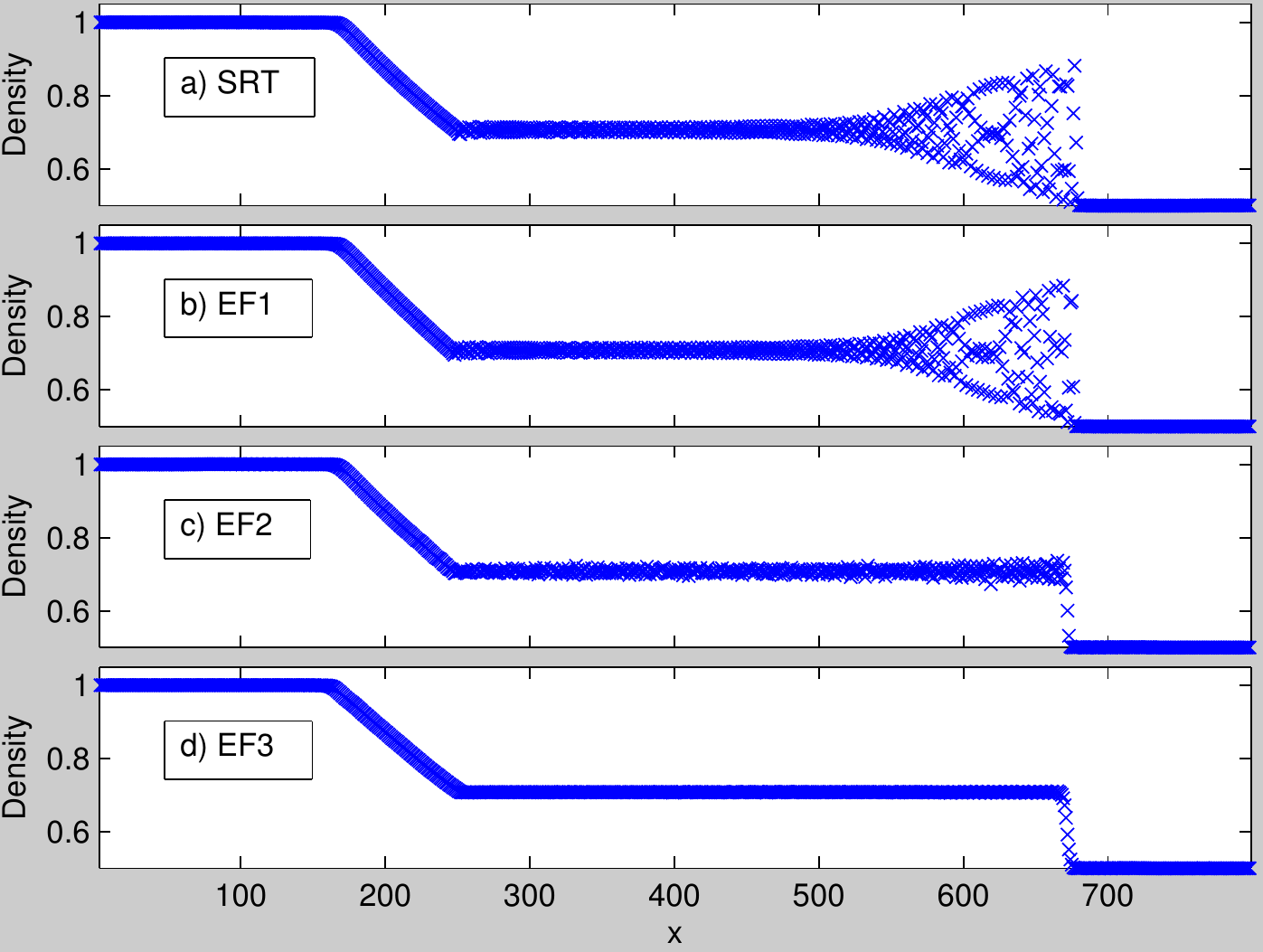}

        \caption{1D shock tube density profiles after 400 times steps. a) SRT-LBM, b) EF1-LBM, c) EF2-LBM, d) EF3-LBM }\label{fig:STLBGKrho}
\end{figure}

\begin{figure}[h!]
        \centering
      \includegraphics[width=.95\columnwidth]{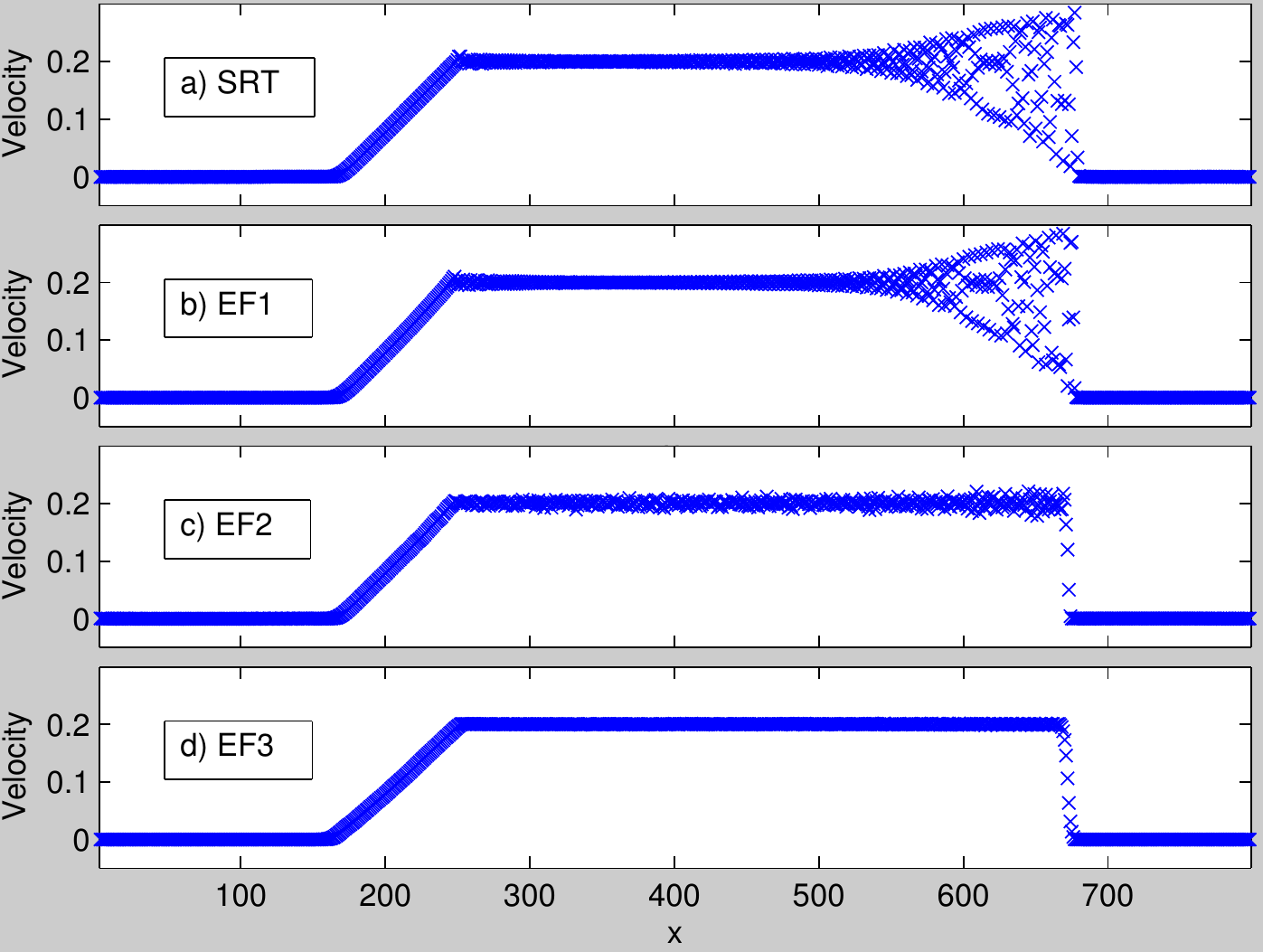}

        \caption{1D shock tube velocity profiles after 400 times steps. a) SRT-LBM, b) EF1-LBM, c) EF2-LBM, d) EF3-LBM }\label{fig:STLBGKux}
\end{figure}

\begin{figure}[h!]
        \centering
      \includegraphics[width=.95\columnwidth]{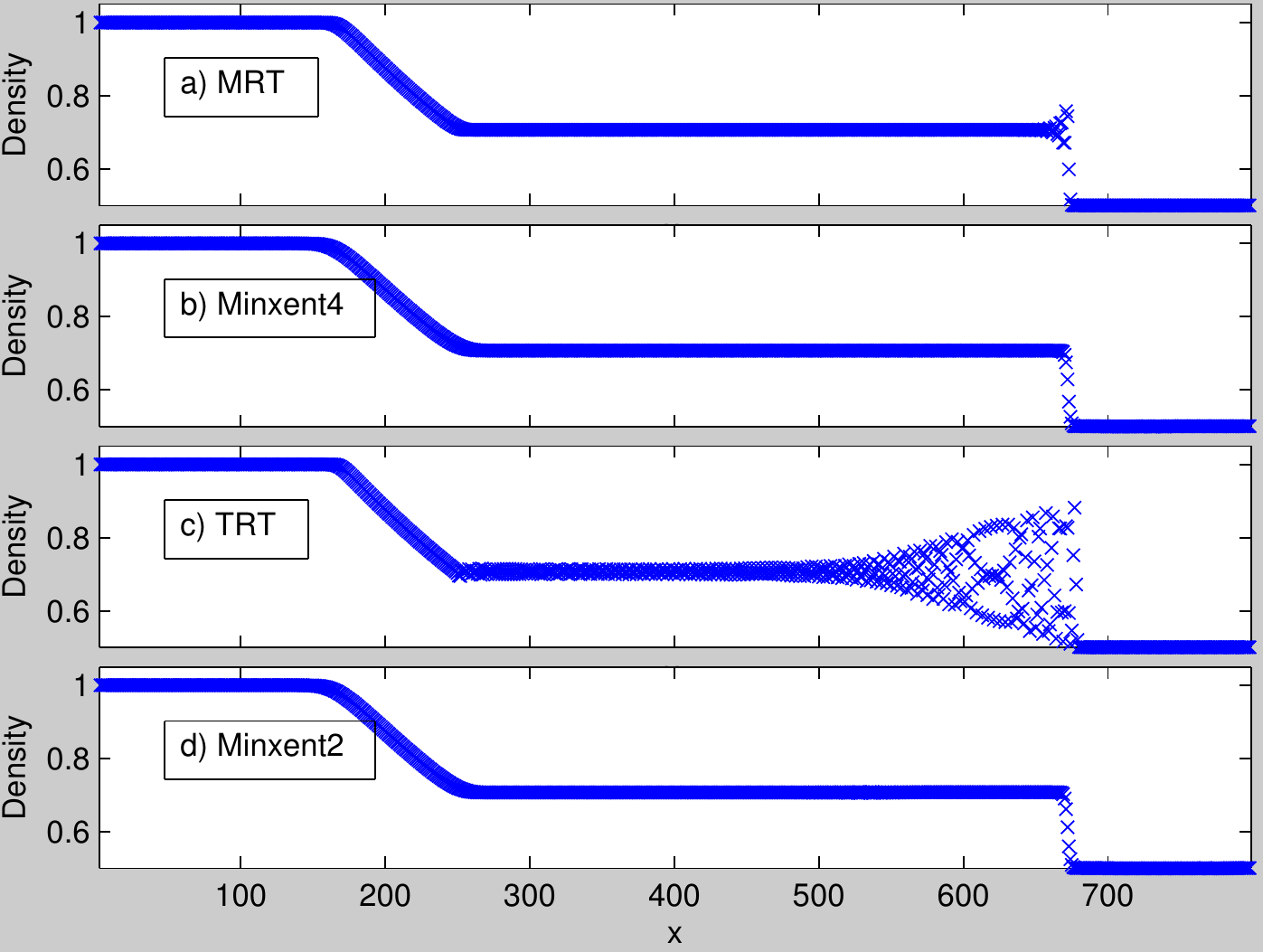}

        \caption{1D shock tube density profiles after 400 times steps. a) MRT-LBM, b) MinxEnt-LBM, c) TRT-LBM, d) MinxEnt2 }\label{fig:STMRTrho}
\end{figure}
\begin{figure}[h!]
        \centering
      \includegraphics[width=.95\columnwidth]{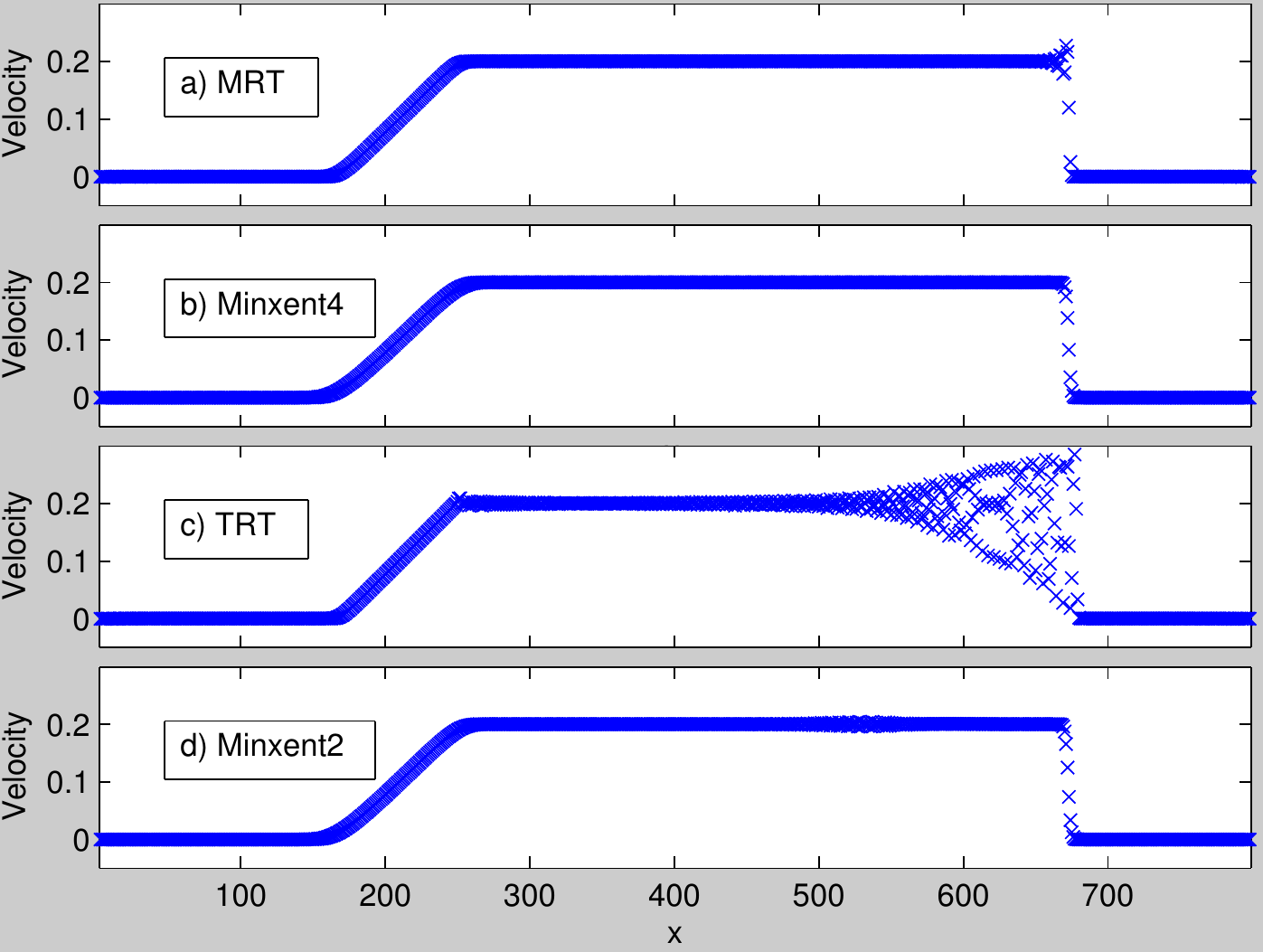}

        \caption{1D shock tube velocity profiles after 400 times steps. a) MRT-LBM, b) MinxEnt-LBM, c) TRT-LBM, d) MinxEnt2 }\label{fig:STMRTux}
\end{figure}
  \subsubsection*{Results}
 
Results for the 1D shock tube are shown in Figures \ref{fig:STLBGKrho}-\ref{fig:STMRTux}. Of particular interest is the behaviour near the shock front. It is clear from the plots that SRT-LBM, EF1-LBM and TRT-LBM suffer from the worst stability, showing severe oscillation near the shock front in both density and velocity. MRT-LBM has improved stability in both density and velocity. The two MinxEnt-LBM based simulations (Figures 3b,3d,4b and 4d) rival the stability of the density and velocity of the lowest tolerance EF-LBM simulations (Figures \ref{fig:STLBGKrho}d and \ref{fig:STLBGKux}d).

\subsection{\label{stabstudies}Lid-Driven Cavity Flow: Stability Studies (LDSS)}

Another benchmark fluid simulation is 2D lid-driven cavity flow. In lid-driven cavity flow the fluid begins at rest and the lid of the cavity is given a constant velocity in the $x$ direction. The remainder of the simulation setup is summarized in second row of Table \ref{table:shocksetup}. A range of $\tau$ values is considered including values approaching zero viscosity ($\tau \to 0.5$). For each value of $\tau$, simulations were carried out with decreasing lid velocities. Simulations are considered ``stable'' the distribution populations remained finite and non-negative at every lattice node and each of the first 1000 time steps.  A  distribution population that has negative components frequently precedes instabilities. This is the reason we choose to label simulations with such distribution populations as ``unstable".  The maximum lid velocity at which a simulation is stable is noted for each value of $\tau$.

 \subsubsection*{Results}
\begin{figure}[h!]
        \centering
      \includegraphics[width=\columnwidth]{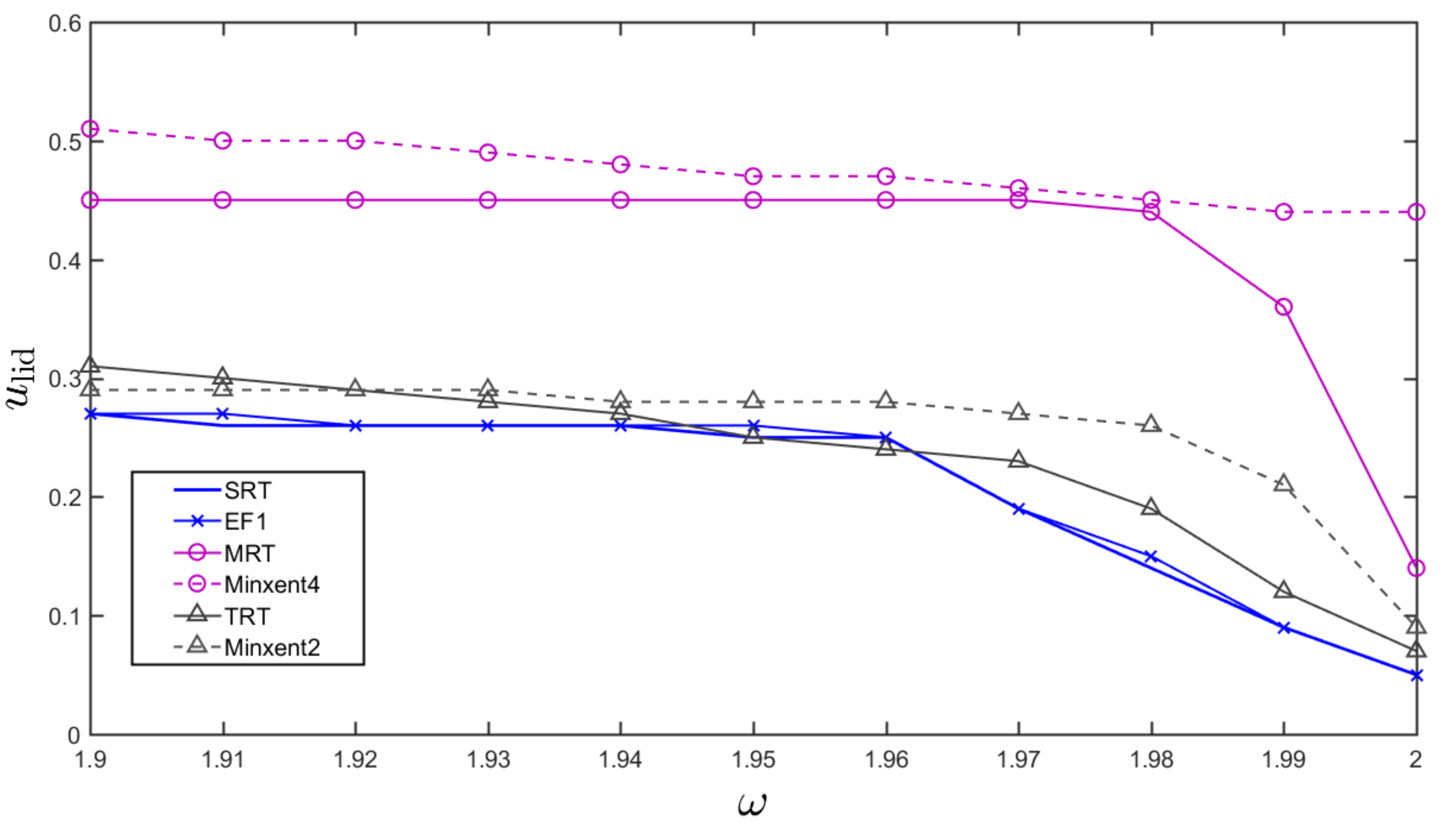}

        \caption{Maximum lid velocity in lid-driven cavity flow allowing simulation to survive 1000 time steps for various viscosity relaxation times. }\label{fig:stabilityplot}
\end{figure}

Results of the lid-driven cavity flow stability simulations are shown in Figure \ref{fig:stabilityplot}. The best performing methods are the MRT-LBM and MinxEnt4 with similar stability, with MinxEnt4 consistently faring slightly better than MRT-LBM. The next best performers were TRT-LBM, MinxEnt2; SRT-LBM the worst. MinxEnt2 is consistently somewhat more stable than its counterpart, TRT-LBM. EF3 and EF2 simulations (not shown) were stable at all lid velocities below 1, and for all values of $1/\tau$ between $1.9$ and $2$. Velocities of 1 or larger were not considered because of the form of \Ref{numerical-enteqm}. 

\subsection{\label{lidstudies}Lid-Driven Cavity Flow: Accuracy Studies (LDAS)}
To assess accuracy, another version of  lid-driven flow simulations were carried out and compared to results from commercial CFD software, COMSOL. These simulations were similar to other studies such as SRT-LBM lid-driven flow (examined by Hou et al. in \cite{Hou1995}) and also MRT-LBM lid-driven flow (by Luo et al. \cite{Luo2011}). In addition to these, Brownlee and co-workers \cite{Brownlee2011} studied lid-driven cavity flows at various Reynolds numbers and using various LBM stabilization techniques.  

In each simulation, a lattice was constructed with $257^2$ nodes ($65^2$ and $129^2$ simulations were also carried out, see \cite{Wilson2016}). A velocity was imparted on the top of the cavity in the $x$ direction.

To calculate the stream function,
$$\psi = - \int u_y({\bf x},t)\, dx $$
we used Simpson's rule for where the lattice was uniformly spaced and the trapezoidal rule where it was not. We used the same formulas for vorticity and normalization of results as in \cite{Luo2011}. 

To standardize the results, the stream function was normalized to the lid velocity, the pressure deviation was normalized to the square of the lid velocity, and the vorticity was normalized to the lid velocity,
$$\hat{\psi}= \frac{\psi}{|{\bf u}_{\rm lid}|}, \qquad \hat{\delta \pi} = \frac{\delta \pi}{{\bf u}_{\rm lid}^2},\qquad \hat{\omega}=\frac{\omega}{|{\bf u}_{\rm lid}|}.$$

\begin{table}[h!]
\small
\centering
\caption{Main Vortex Results, $N_x,N_y=257$.  Top Row: Main Vortex, Middle Row: Lower Right Vortex, Bottom Row: Lower Left Vortex.}
\label{table:results}
    \begin{tabular}{|c|ccccc|}
    \hline
       & $\hat{\psi}_{\rm min}$ & x       & y       & $\hat{\delta \pi}$  & $\hat{\omega}$ \\
    \hline
    Comsol   & -0.11881 & 0.53137 & 0.56445 & -0.074009 & -2.0634 \\
 EF3&-0.13746&0.51569&0.55664&-0.10237&-2.3762\\
   Minxent4   & -0.11808 & 0.53137 & 0.56445 & -0.073515 & -2.0552 \\
  \hline
    Comsol    & 1.7192 & 0.86471 & 0.11133  & 3.5281   & 1.0996  \\
   EF3&1.785&0.86471&0.12695&4.3053&1.2018\\
   Minxent4    & 1.7071 & 0.86471 & 0.11133  & 3.4503   & 1.0801  \\
    \hline
    Comsol    & 2.2514  & 0.084314 & 0.076172 & 4.4704 & 3.5089  \\
   EF3&2.9686&0.088235&0.076172&5.5588&4.2134\\
   Minxent4    & 2.255   & 0.084314 & 0.076172 & 4.382  & 3.4119  \\
    \hline
    \end{tabular}
\end{table}

\subsubsection*{Results}
\begin{figure}[h!]
        \centering
      \includegraphics[width=0.9\columnwidth]{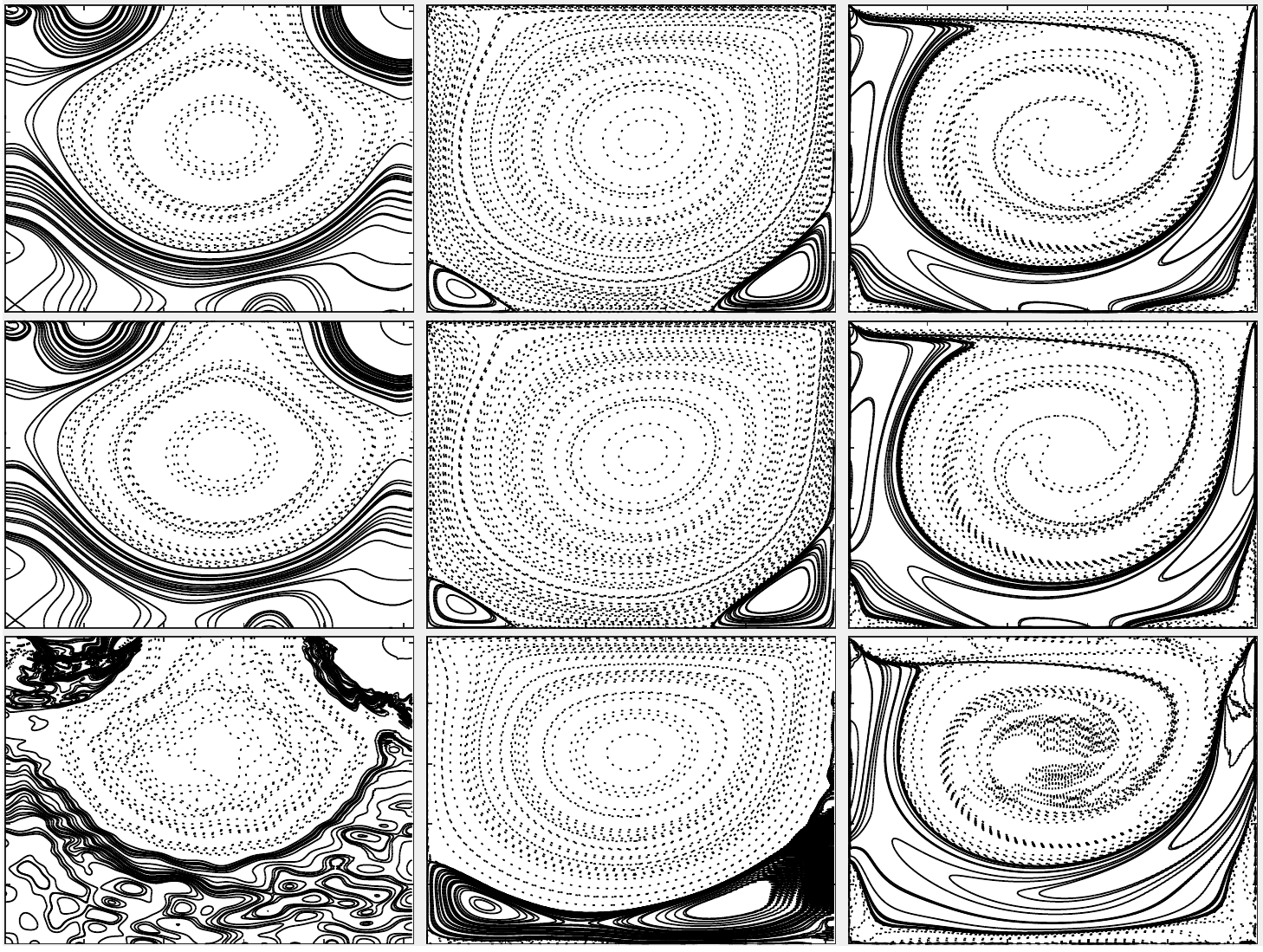}

        \caption{Flow contours of lid-driven cavity flow for Re$=1000$ with $N=257^2$. Left: Pressure deviation, Middle: Stream Function, Right: Vorticity. Top Row: Comsol, Middle Row: SRT-LBM, Bottom Row: EF3-LBM}\label{fig:LID1000}
\end{figure}

The results of the $257 \times 257$ node lid-driven cavity flows for $Re=1000$ were the most visually interesting and are the only results reported here; see Figure \ref{fig:LID1000}. Numerical results are given in Table \ref{table:results}. Presented in the table are the centres (determined by the extrema of the streamfunction) of the main, lower-left, and lower-right vortices. In addition, the pressure deviation  and vorticity at these locations is presented. For simulations with $65^2$ and $129^2$ nodes as well as for smaller Reynolds numbers, see \cite{Wilson2016}. Also refer to \cite{Wilson2016} for simulations using other collision rules discussed in this manuscript.

\subsection{Discussion}
To explore the the effect on stability of different collision rules, two types of simulations were conducted. In the 1D shock tube simulations, we can see from Figures 1-4 that, although MRT-LBM shows improved stability over SRT-LBM, the only collision rules that maintained a sharp shock front were EF3-LBM, Minxent4 and Minxent2.  All other collision rules experienced instability at the shock front. This seems to indicate that EF3-LBM, Minxent4 and Minxent2 are candidates for the most stable collision rule. 

Similar conclusions are found in results of the lid driven stability simulations, shown in Figure 5. Though not plotted in Figure 5, simulations employing EF2-LBM and EF3-LBM collision rules remained stable for all lid velocities below 1 and for values of $1/\tau$ between $1.9$ and $2$. Simulations were not carried out for EF3-LBM and EF2-LBM at a lid velocity of 1 owing to the form of the equilibrium \Ref{numerical-enteqm}. It is unsurprising that EF2-LBM and EF3-LBM are able to survive at all tested lid velocities because when $\tau$ is close to $0.5$, the effect of the EF2-LBM and EF3-LBM collision rules \Ref{EFrule} is to return the distribution to near-equilibrium whenever $\delta S$  exceeds tolerance. Effectively, whenever a lattice node was in danger of losing stability (indicated by $\delta S$ above tolerance), the distribution was set to equilibrium. This ensured the simulation always remained stable. Other than EF2-LBM and EF3-LBM, the collision rule that survives at the highest lid velocities is Minxent4 followed by MRT, Minxent2 and TRT. The least stable collision rule was SRT-LBM. From these observations we can further conclude that the collision rules that lead to the most stable simulations are EF3-LBM and Minxent4. 

We next tested the accuracy of the these two, most stable, collision rules. To accomplish this we performed lid driven cavity flow simulations for various Reynolds numbers and compared the results to simulations using commercially available software (Comsol) under the same flow conditions. From Figure \Ref{fig:LID1000} we see that EF3-LBM deviates considerably from the Comsol results. As shown, the Minxent4 simulations qualitatively reproduced the Comsol results. These conclusions are also quantitatively supported in Table \ref{table:results} where we see that results using the Minxent4 collision rule are much more similar to Comsol than the results using the EF3-LBM collision rule. 

From these stability and accuracy tests we find that Minxent4 offers the best mix of stability and accuracy in the simulations discussed here. 

It is worth mentioning that improved accuracy for EF2-LBM and EF3-LBM collision rules have been reported in the literature \cite{Brownlee2006,Brownlee2007}. However, in order to maintain accuracy, these studies limited the number of lattice nodes where the ``more gentle" collision rule  \Ref{gentle-ef-rule} was used. That is, using the EF-LBM collision rules, there is a trade-off between accuracy and stability. This trade-off is mediated by the maximum number of lattice nodes permitted to use \Ref{gentle-ef-rule}. The more lattice nodes that are allowed to use \Ref{gentle-ef-rule}, the more stable the simulation is, but the less accurate the simulation. The fewer lattice nodes that use \Ref{gentle-ef-rule}, the more accurate the simulation is but the less stable it is. This means, for EF-LBM, the tolerance and maximum number of lattice nodes permitted to use \Ref{gentle-ef-rule}, are parameters that need to be tuned.

Likewise, MRT and TRT both show improved stability over SRT, but they are both dependent on the choice of relaxation times that are not related to the fluid viscosity. Thus, similar to EF-LBM, MRT and TRT simulations can exhibit improved stability, but require parameters to be tuned. In MRT-LBM, relaxation times associated with the second and third moments could be adjusted to increase stability. In TRT-LBM $\tau_2$ could be adjusted rather than use its prescribed value given in \Ref{tau2}. 

Tuning these parameters would be need to be performed on a simulation by simulation basis. Unlike the other collision rules simulated here, This is contrasted with Minxent-LBM methods which  able to improve stability, without losing accuracy and without needing to tune any parameters.

\section{\label{conclusions}Conclusions}
In this paper we have derived a novel collision step for the Lattice Boltzmann Method based on the Principle of Minimum Cross Entropy, MinxEnt-LBM. 

MinxEnt-LBM was used in numerical simulations and compared to existing LBMs. The only scheme that showed comparable stability to MinxEnt-LBM was the entropy limiting scheme of the LBM based on Ehrenfest Steps (EF-LBM). However, lid driven cavity flow simulations showed that without tuning the parameters involved in EF-LBM schemes they suffered from a degradation of accuracy. We can conclude that, of the LBM schemes tested, MinxEnt-LBM had the best combination of stability and accuracy.

An important practical consideration is that EF-LBM and MRT-LBM require specification of parameters which need to be tuned and optimal values are not known {\it a priori}. This is not the case for the MinxEnt-LBM.

\bibliography{Mendeley.bib}
\end{document}